\documentclass[preprint,journal]{vgtc}            

\onlineid{8605}
\preprinttext{}

\vgtccategory{Research}

\usepackage[]{cite} 
\usepackage{amsmath,amsfonts}
\usepackage{graphicx}
\usepackage[svgnames]{xcolor}
\usepackage[most]{tcolorbox}
\usepackage{array}
\usepackage{float}
\usepackage[caption=false,font=normalsize,labelfont=sf,textfont=sf]{subfig}
\usepackage{textcomp}
\usepackage{csquotes}
\usepackage{stfloats}
\usepackage{url}
\usepackage{booktabs}
\usepackage{tikz}
\usepackage{framed}
\usepackage{amsthm}
\usepackage{wasysym}
\usepackage{soul}
\usepackage{enumitem}
\usepackage{multirow}
\usepackage{multicol}
\usepackage{ifthen}
\usepackage{svg}
\usepackage{placeins}
\usepackage[english]{babel}
\usepackage[subtle]{savetrees}
 
\usepackage[pagebackref, bookmarks]{hyperref}
\addto\extrasenglish{

}
	
\newcommand{\orcid}[1]{\hspace*{0.2em}\hbox{\href{https://orcid.org/#1}{\includegraphics{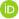}}}}

\newcommand{\revise}[1]{{#1}}

\newcommand{\subhead}[1]{\vspace{2pt} \noindent \textbf{#1 --}}

\definecolor{c_ex}{RGB}{255, 149, 0} 
\definecolor{c_fl}{RGB}{50, 173, 230} 
\definecolor{c_gt}{RGB}{52, 199, 89} 
\definecolor{c_ct}{RGB}{255, 45, 85} 

\newcommand{\stcOperation}[2]{%
    \begingroup
    \def\tempcolor{}%
    \ifthenelse{\equal{#1}{ex}}{\def\tempcolor{c_ex!75}}{%
    \ifthenelse{\equal{#1}{fl}}{\def\tempcolor{c_fl!75}}{%
    \ifthenelse{\equal{#1}{gt}}{\def\tempcolor{c_gt!75}}{%
    \ifthenelse{\equal{#1}{ct}}{\def\tempcolor{c_ct!75}}{%
    \def\tempcolor{black}}}}}
    \tikz[baseline=(X.base)] \node[
        fill=none,
        draw=\tempcolor,
        rounded corners=2pt, 
        inner xsep=2pt, 
        inner ysep=0
    ] (X) {\strut \textbf{#2}};%
    \endgroup
}

\newcommand{\toolName}{STcubeOperator}
\newcommand{\papertitle}{\toolName: A Framework for Analyzing Spatiotemporal Event Data}

\newcommand{\authorsshort}{
	{Julius~Rauscher}\orcid{0000-0001-7049-5203},
	{Lucas~Joos},\orcid{0000-0001-7049-5203},
    {Mark Rohonyi},
	{Daniel~A.~Keim}\orcid{0000-0001-7966-9740},
	and {Maximilian~T.~Fischer}\orcid{0000-0001-8076-1376}
	}
\newcommand{\authorsfull}{
	All authors are with the University of Konstanz. E-mail: \{julius.rauscher, lucas.joos, mark.rohonyi, keim, max.fischer\}@uni-konstanz.de.
}

\newcommand{\keywordlist}{Space-Time Cubes, Spatiotemporal Visualization, Event Data, Geospatial Analysis, Visual Analytics.}

\newcommand\figCapVSpace{-1.4em}
\newcommand\figVSpace{-1.2em}

\usepackage{titlesec}
\titlespacing*{\subsection}
  {0pt}{*0.8}{-2pt}
\titlespacing*{\subsubsection}
  {0pt}{*0.6}{-2pt}

\newcommand{\usr}[1]{\textbf{E\textsubscript{#1}}}
\newcommand{\tsk}[1]{\textbf{T\textsubscript{#1}}}

\definecolor{c_unident}{rgb}{0.6, 0.788, 0.270}
\definecolor{c_civilians}{rgb}{0.898, 0.525, 0.023}
\definecolor{c_marines}{rgb}{0.364, 0.4117, 0.694}
\definecolor{c_naf}{rgb}{0.321, 0.737, 0.639}

\definecolor{c_air}{rgb}{0.898, 0.525, 0.023}
\definecolor{c_ground}{rgb}{0.364, 0.4117, 0.694}

\newcommand{\colorrect}[1]{\textcolor{#1}{\rule{.75em}{.75em}}\hspace{.1em}}

\title{\papertitle}

\author{\authorsshort}

\authorfooter{\authorsfull}

\abstract{%
    The analysis of spatiotemporal event data is essential for informed decision-making in domains such as disaster response, conflict analysis, or intelligence investigations. 
    However, the complexity and interdependence of spatial, temporal, and multiple thematic attributes pose significant challenges for both analysis and visualization.
    While space-time cubes (STCs) present a powerful integrated visualization technique to analyze this kind of data, existing approaches often lack support for complex exploratory workflows, 
    thus limiting the ability to derive meaningful insights.
    We address this gap by introducing \toolName, a novel framework that models analysis tasks through space-time cube operations, considering them in context of visualizations, interactions, and computational choices, and implement them in an interactive visual analytics environment.
    By expressing analysis tasks as a sequence of multiple elementary operations--such as filtering, chopping, and flattening--our approach enables analysts to dynamically explore data from different perspectives.
    We further provide an open-source prototype implementing the operations in a 3D interactive environment to facilitate task-based exploratory analysis of spatiotemporal event data.
    We demonstrate the applicability of our framework with a case study based on real-world data on strategic and military operations in the Russia-Ukrainian War, showing its capabilities to reveal spatiotemporal patterns.
    An expert user study (n=8) shows how specific tasks can be solved with our framework, highlights the versatility of our approach, and provides valuable insights on which operations experienced analysts utilize in practice.
}


\teaser{
  \centering
  \includegraphics[width=\linewidth]{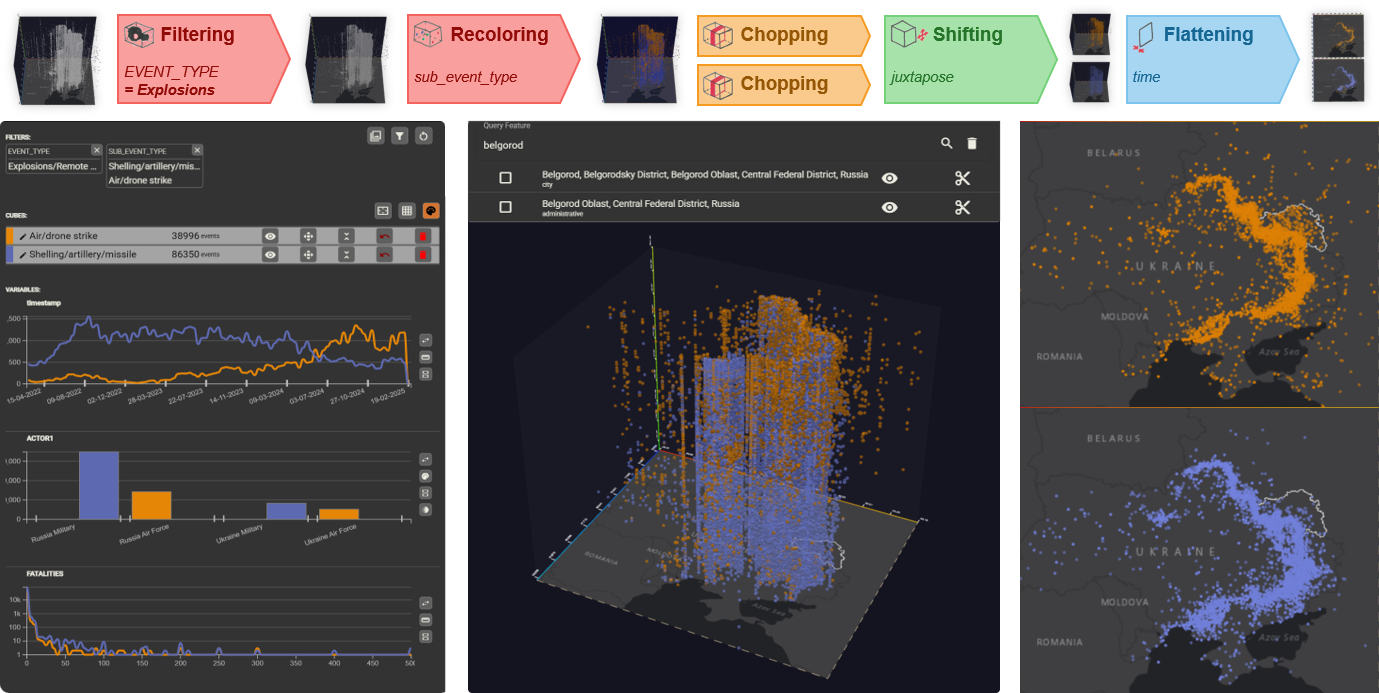}
  \caption{%
One workflow of using \toolName~to perform \textbf{comparative geospatial analysis of the Russia-Ukrainian War} (data from February 2022 to February 2025), specifically weapon usage over time. After filtering the data, a recoloring is applied discerning \colorrect{c_air}\emph{Air/drone Strikes} from \colorrect{c_ground}\emph{Shelling/Artillery} events. For a better comparison, the cube is repeatedly chopped using categories of the \emph{sub\_event\_type} and shifted side by side followed by a flattening operation (right).
While the cubes reveal little spatial deviation in general, a distinction can be identified in the region of Belgorod.
The STC (center, space-time view) visually indicates a temporal shift from \emph{Shelling/Artillery} to \textit{Air/drone Strike} events over time, which is confirmed by the temporal distribution plot (left, attribute views).
  }
  \label{fig:teaser}
}

\graphicspath{{figs/}{figures/}{pictures/}{images/}{./}} 

\begin{document}


\firstsection{Introduction}

\maketitle

The analysis of spatiotemporal event data plays a crucial role in various domains, including disaster response~\cite{Gatalski.SpaceTimeCube.2004, Thom.stAnomalyTwitter.2012}, armed conflicts~\cite{Weidemann.WarViews.2009, swedberg.vaStCalendar.2017}, crime analysis~\cite{abusalah.vaHotspots.2010, wang.crimedistva.2013} or traffic analysis~\cite{Deng.VisCas.2018, Wagner.TaxiVisVR.2024}, serving as a foundation for informed decision-making.
The complex interplay of spatial, temporal, and additional thematic attributes imposes a core challenge for visualization and analysis~\cite{hamdi.stDataMining.2022}, \revise{as effective visual encodings must reconcile the high dimensionality of the data with the limited representational screen space, which is often two-dimensional.}

To conceptualize the underlying analysis problem, task taxonomies have been proposed~\cite{Amar.lowlvTasks.2005,Andrienko.ExploratoryAnalysis.2006, brmer.vistasks.2013}, facilitating a concrete formalization of exploratory objectives.
By adopting a structured task-based approach, questions can guide the visual exploratory analysis process \revise{and help to reduce the dimensional complexity, directing attention towards more relevant attributes.}

While 2D visualization approaches~\cite{ferreira.taxivis.2013,Deng.VisCas.2018, silva.VAstEvents.2019} can support task-based analysis workflows, they are limited in their ability to represent complex relationships among spatial, temporal, and thematic attributes within an integrated view.
3D visualizations offer an enhanced capacity to encode an additional dimension, providing greater flexibility to represent multidimensional data aspects. 
The space-time cube (STC) is a well-established visualization technique~\cite{Gatalski.SpaceTimeCube.2004, Andrienko.ClusterAnalysisSTC.2009, Kveladze.GVAtool.2015} that extends traditional 2D representations by introducing a third dimension for time, allowing analysts to observe spatiotemporal trends and effectively support them in solving analysis tasks~\cite{Kristensson.CompareSpaceTimeCubeTo2D.2009, Kjellin.2Dvs3D.2010, Kveladze.GVAtool.2015} more efficiently than in 2D.
Despite the known shortcomings of 3D environment tools (i.e. occlusion, perspective distortion, depth ambiguity), various studies have demonstrated this effectiveness and efficiency of STCs, indicating advantages over 2D visualizations for observing complex spatiotemporal
patterns~\cite{Kristensson.CompareSpaceTimeCubeTo2D.2009}, retrieving temporal ordinal information~\cite{Kjellin.2Dvs3D.2010}, or examining event sequences~\cite{Amini.InteractivityImpact2Dvs3D.2015}.
\revise{However, existing implementations of STCs are not openly available, and} although interactivity is a key component~\cite{Kapler.GeoTime.2004, Amini.InteractivityImpact2Dvs3D.2015}, they often lack flexibility and are tailored towards insular aspects, \revise{such as clustering~\cite{Andrienko.ClusterAnalysisSTC.2009, Nakaya.StcCrimeClusters.2010}, sequential patterns~\cite{Yusof.WindpatternsSTC.2016}, or clutter reduction~\cite{Demsar.SpaceTimeDensity.2010}}, falling short in addressing a broad range of complex analysis tasks. 

The concept of a generalized STC has been introduced by Bach et al.~\cite{Bach.SurveySpaceTimeCubeOperations.2016} as a descriptive model to categorize temporal data visualizations, where existing visualizations can be described as operations on a conceptual cube.
As such, the famous Minard map~\cite{tufte.visualDisplayMinard.1983} is an example of a Time Flattening operation, combining spatial and temporal information on a single 2D canvas.
Depicting temporal information along a curvilinear space path as a 2D plot (i.e. Mareys Train schedule~\cite{marey.trainschedule.1878}) can be seen as a Space Cutting operation. 
Bach et al.'s framework provides an exhaustive taxonomy of \emph{operations} that is purely conceptual. 
While these theory- and conceptual-driven taxonomies of STC operations have been utilized to structure, categorize, and describe existing visualizations, they ideally should be available GIS users~\cite{Andrienko.SpaceTimeVA.2010}, but so far they have never been practically implemented as an interactive framework. 
\revise{To fill this gap, we introduce a practical framework that operationalizes the conceptual taxonomy of STC operations and} facilitates a structured, task-based exploration by expressing tasks as a sequence of operations.

To summarize, existing STC approaches are often tailored to single use cases and are not openly accessible, and interactive tools implementing STC operations provide only limited operational coverage.
We fill this gap by exploring how the analysis process and corresponding use cases can be described by elementary and compound operations.
We explore task-operation mappings, implement these operations in a visual analytics (VA) application, and evaluate the applicability in both a comprehensive case study and expert study.
Thereby, we make the following contributions:

\begin{itemize}[topsep=0pt,itemsep=-.75ex,partopsep=1ex,parsep=1ex, leftmargin=1em]
    \item \toolName~framework, deconstructing theoretical STC operations towards visualization, interaction, and computation.

    \item An \href{https://osf.io/qgr4x/overview?view_only=48a6e22e60574460a47ef24102f3c78a}{open-source} implementation as part of a \href{https://stcubeoperator.dbvis.de/}{online \emph{VA prototype}}.

    \item An evaluation of the framework through both a \emph{case study} on strategic movements in the Russo-Ukrainian war and an \emph{expert user study} demonstrating task coverage and operation sequences.
    
\end{itemize}

\section{Related Work}
\label{sec:related_work}

Our work relates to the fields of (1) spatiotemporal event visualization, (2) STCs, and the concept of (3) space-time operations. 

\subsection{Spatiotemporal Event Visualizations}

Spatiotemporal data and VA analysis techniques have been described exhaustively by Andrienko and Andrienko, emphasizing the necessity for visual tools consisting of multiple linked views to uncover patterns~\cite{Andrienko.ExploratoryAnalysis.2006, Andrienko.VAMovementSpatialEvents.2013}.
GIS applications such as QGIS~\cite{Moyroud.QGIS.2018}, ArcGIS~\cite{Scott.ArcGIS.2010}, or kepler.gl~\cite{kepler.Uber.2018} provide flexible data visualization interfaces targeted at a broad user group, yet their standardized methods can fall short in creating custom solutions or require specialized, often incompatible plugins.

One commonly used visualization technique are thematic maps, typically coupled with linked temporal views~\cite{Ferreira.BirdVis.2011} or overlaid embeddings~\cite{Sun.MapRouteZooming.2017} that further allow insights into temporal dynamics. 
Other forms to encode temporal information include Glyph Maps~\cite{Beecham.Glyphmaps.2021} or Small Multiples~\cite{Johnson.InternetSmallMultiples.2015}. 
To reduce the complexity of the data, the geography can also be represented in abstract ways~\cite{Rauscher.1dorder.2025}, through pixel-based layouts~\cite{Andrienko.MobilityAbstraction.2017} or dimensionality reduction methods~\cite{Zhou.ODAbstraction.2019}.
VA frameworks often integrate multiple linked views to enhance event analysis.
Hotspot analysis, including kernel density estimation maps~\cite{abusalah.vaHotspots.2010} and heatmap visualizations~\cite{wang.crimedistva.2013} highlight spatial concentrations.
Social media data has been leveraged to develop topic maps~\cite{Thom.stAnomalyTwitter.2012} for anomaly detection.
Task-based exploration is supported through visual query models~\cite{ferreira.taxivis.2013}, while calendar-based views~\cite{swedberg.vaStCalendar.2017} can expose recurring temporal patterns.
Cascades~\cite{Deng.VisCas.2018} can show coherent event sequences, and adaptive level-of-detail approaches~\cite{silva.VAstEvents.2019} permit analysis across various granularities.

Despite the diversity of existing 2D VA approaches, space and time are typically analyzed in separate components, such as maps for spatial representation and timelines for temporal trends. This separation introduces a visual discontinuity that disrupts spatiotemporal pattern discovery.
STC visualizations~\cite{Hägerstrand.SpaceTimePaths.1970} present a promising solution, allowing a unified visual representation.

\subsection{Space-Time Cube Visualizations}
In 1970, Hägerstrand~\cite{Hägerstrand.SpaceTimePaths.1970} laid the conceptual foundation for STCs when he introduced the space-time concept, representing lifetimes as a path through space and time.
This concept was later adopted by Carpendale et al.~\cite{Sheelagh.Tardis.99} to analyze temporal dynamics of spatial landscape patterns by means of spatiotemporal blocks.
Various versions of an STC framework have been proposed since, featuring interactivity~\cite{Kraak.SpaceTimeCubeGeoPerspective.2003}, linked time charts~\cite{Kapler.GeoTime.2004}, 
integrated timelines~\cite{Yusof.WindpatternsSTC.2016},
filtering, scaling, coloring cube objects according to additional attributes~\cite{Gatalski.SpaceTimeCube.2004}, clustering functionalities~\cite{Andrienko.ClusterAnalysisSTC.2009}, 
or density volumes for clutter reduction~\cite{Demsar.SpaceTimeDensity.2010}.
In a survey and outlook on spatiotemporal visualization, Zhong et al.~\cite{Zhong.SpatiotemporalVisSurvey.2012} describe possible operations that could be applied to STCs as future work.
Kraak~\cite{Kraak.SpaceTimeCubeGeoPerspective.2003} proposed an interactive STC and highlighted the potential of coupled linked views.
With GeoTime~\cite{Kapler.GeoTime.2004}, Kapler and Wright developed an interactive STC framework that further integrates visual time-tracks.
Gatalski et al.~\cite{Gatalski.SpaceTimeCube.2004} discuss the option of scaling and coloring objects within the cube according to additional attributes and provide filtering to mitigate overplotting and occlusion problems.
Andrienko and Andrienko~\cite{Andrienko.ClusterAnalysisSTC.2009} integrate a clustering approach for STCs.
Demšar and Virrantaus~\cite{Demsar.SpaceTimeDensity.2010} introduce the concept of space-time density to reduce visual clutter in STCs, \revise{while Nakaya and Yano~\cite{Nakaya.StcCrimeClusters.2010} further propose using colored cylinders to encode statistical significances of space-time clusters}. 
Kveladze et al.~\cite{Kveladze.stc_vis_variables.2019} investigate the mapping of visual variables in STCs.
%
\revise{Deng et al.~\cite{stcVolumes:deng:2025} propose an STC-based implementation to explore patterns in spatial timeseries data through volume and surface renderings. }
More recent studies investigate an immersive STC setup with virtual table tops~\cite{Zhang.TimeTables.2022}, gesture interaction~\cite{Filho.ImmersiveTableSpaceTimeCube.2020, Wagner.TaxiVisVR.2024}, or extensions to a Hypercube for biomedical 3D spatial data analysis~\cite{Fouche.ImmersiveHyperSpaceTimeCube.2023}.
%
Other 3D-based approaches to visualize spatiotemporal data include pencil or helix icons~\cite{Tominski.3DPencils.2005, Gautier.covidCube.2021}, 3D shapes~\cite{Forlines.Wakame.2010}, a space-time wall~\cite{Goesele.STWall.2012}, \revise{as well as temporal transformations of the wall~\cite{trajectoryWall:andrienko:2014}}.

To evaluate the effectiveness and efficiency of STCs, various task-based studies have been conducted, comparing the system to a 2D map representation.
Kristensson et al.~\cite{Kristensson.CompareSpaceTimeCubeTo2D.2009} compared a STC visualization to a baseline 2D map representation, indicating a higher error rate for simpler tasks, but a lower response time for observing complex patterns. 
Kjellin et al.~\cite{Kjellin.2Dvs3D.2010} concluded that STCs fall short when 
metric properties are crucial, but outperform 2D-based approaches when the order of arrival of moving objects should be determined.
Willems et al.~\cite{Willems.STCvsDensityMap.2011} investigate
how different visualizations (moving dot maps, STCs, and vessel density) capture various movement features (Lanes, Stops, Fast).
According to their study, STCs perform best in identifying lanes, while exhibiting a large spread among participants regarding correctness and response time for identifying stops.
Kveladze et al.~\cite{Kveladze.GVAtool.2015} proposed a GeoVisual analytics environment consisting of a 2D map, a STC, and a Parallel Coordinate Plot.
Their results suggest that the STC is imperative for identifying temporal patterns, and some participants even mentioned they could solve all tasks with this view alone.
Amini et al.~\cite{Amini.InteractivityImpact2Dvs3D.2015} conducted an experimental comparison between a STC and a 2D map regarding a set of questions on movement data.
The participants consistently preferred the STC over 2D visualizations for all questions. 
Especially when sequences of events had to be examined, the STC was determined as beneficial.

\revise{Despite these well-established academic research descriptions and evaluations, variants have also been implemented in commercial applications like ArcGis~\cite{Scott.ArcGIS.2010}. 
}
Yet, openly accessible applications implementing this concept remain rare, and existing implementations lack standardized operations that comprehensively support a broad range of diverse analysis tasks.

\subsection{Space-Time Cube Operations}
The concept of STC operations is present in the literature as an abstract representation form. 
Bach et al.~\cite{Bach.SurveySpaceTimeCubeOperations.2016} present a descriptive model to categorize visualizations of temporal data based on a set of STC operations. 
While their model is based on a STC, it is purely conceptual and does not necessitate a geographical component.
Li et al.~\cite{Li.TextSpaceTime.2020} leverage a STC metaphor for systematic text analysis, and define various slicing and aggregating operations that can be utilized to investigate the spatiotemporal development topics within social media. 
Bayomie et al.~\cite{Bayome.STCinProcessMining.2020} refer to the STC operations defined by Bach et al.~\cite{Bach.SurveySpaceTimeCubeOperations.2016} to categorize tools in the domain of process mining.

While these approaches define STC operations mainly for categorization purposes, they are treated as a conceptual framework, and have neither been \revise{realized in a unified implementation} for practical visual data analysis nor evaluated in a real-world context.

\section{The \toolName~Framework}
\label{sec:methodology}

\begin{table*}[t]
    \centering
    \renewcommand{\arraystretch}{0.9}
    \small
    \resizebox{\textwidth}{!}{
    \begin{tabular}{@{}lllll@{}}
        \toprule
        \multirow{2}{*}{\textbf{ID}}  
            & \textbf{Elementary /} & \textbf{Direct /} 
            & \multirow{2}{*}{\textbf{Task Type}}  
            & \multirow{2}{*}{\textbf{Examplary question used in Expert Study}} \\
            & \textbf{Synoptic} & \textbf{Inverse} &   &   \\
        \midrule
        \tsk{1}  & elementary & direct & Lookup        & How many \textit{fatalities} were recorded at the Music Festival on \textit{October 7th, 2023} in \textit{Reim}? \\
        \tsk{2}  & elementary & inverse & Lookup        & When and Where did \textit{Non-state actor overtakes territory} occur? \\
        \tsk{3}  & synoptic & direct & Lookup        & How was the \textit{Violence against Civilians} of the \textit{Hamas Movement} on \textit{October 7th, 2023} spatially distributed? \\
        \tsk{4}  & synoptic & inverse & Lookup        & Find a ceasefire interval (no \textit{explosions}) after the invasion on \textit{October 7th, 2023}. \\
        \tsk{5}  & elementary & direct & Comparison    & Compare the events where \textit{Sexual Violence} was recorded. In which dimensions do they differ? \\
        \tsk{6}  & elementary & inverse & Comparison    & Compare the two events with the highest \textit{fatalities}. When? Where? What actors? What event types? \\
        \tsk{7}  & synoptic   & direct & Comparison    & Are more \textit{Protests} happening inside of \textit{Israel} or outside? \\
        \tsk{8}  & synoptic & inverse & Comparison    & Compare the spatial distribution of the actors \textit{Lions’ Den} and \textit{Al Aqsa Martyrs Brigade}. \\
        \tsk{9}  & elementary   & \emph{n/a} & Rel.-Seeking     & Find days where there were more \textit{Protests} than \textit{Riots}. \\
        \tsk{10} & synoptic & \emph{n/a} & Rel.-Seeking     & At which time did the number of \textit{Explosion} events in Lebanon exceed the number of \textit{Explosion} events in Gaza?\\
        \bottomrule
    \end{tabular}
    }
    \caption{Analysis tasks (adapted from~\cite{Andrienko.ExploratoryAnalysis.2006}) and example questions used in the expert study (see~\autoref{sec:userstudy}).}
    \label{tab:tasks}
    \vspace{\figVSpace}
\end{table*}

The literature shows that STCs can be an effective way to visualize and analyze spatiotemporal event data, \revise{and that STC operations show potential for structured, task-based exploration.
However, commercial~\cite{Scott.ArcGIS.2010, Moyroud.QGIS.2018, kepler.Uber.2018} or scientific~\cite{Andrienko.ClusterAnalysisSTC.2009, Demsar.SpaceTimeDensity.2010, Yusof.WindpatternsSTC.2016}  solutions are not openly accessible, or lack standardized operations to support a broad range of analysis tasks.}
To fill this gap, we propose to express tasks as a \emph{sequence of operations}, in line with compound operations described by Bach et al.~\cite{Bach.SurveySpaceTimeCubeOperations.2016}, that can be interactively performed through a VA framework.
\revise{Bach et al. introduced their generalized STC framework as purely conceptual and} only outlined a possible interactive implementation when comparing it to the InfoVis pipeline~\cite{Card.Vison2Think.1999}.
While the InfoVis pipeline models the transformation of data into a visual format (e.g., a visualization) with interactivity to enable users to modify these transformations, STC operations can be applied at different stages of the pipeline depending on their interpretation~\cite{Bach.SurveySpaceTimeCubeOperations.2016}.
Since exploratory data analysis is often task-driven, we consider analysis tasks a more suitable target for implementing STC operations, as these tasks are often coupled with insights and knowledge generation.
We therefore introduce \textbf{analysis tasks}, \textbf{STC operations}, and their implementation into a VA \textbf{prototype}.
\revise{\autoref{fig:infovis_pipeline} shows their interaction within the \toolName~framework as a \emph{conceptual view}. }
When translating STC operations from a conceptual model into an interactive system, semantic discrepancies can arise due to practical constraints and design choices. 
We discuss these differences and how our implemented versions deviate from their theoretical definitions.


\subsection{Analysis Tasks}
\label{subsec:st_tasks}
A task-based analysis approach offers a clear structure for generating insights and helps to classify how users interact with data, and several taxonomies for analysis tasks have been proposed~\cite{Amar.lowlvTasks.2005, Andrienko.ExploratoryAnalysis.2006, brmer.vistasks.2013}.
A task can be broken down into two key components: the \textbf{target} (\emph{T}), which defines the information to be obtained, and the \textbf{constraints} (\emph{C}), which specify the conditions this information must meet. 
Mathematically, this can be defined as $C \rightarrow T$.
A conceptualization of spatiotemporal dynamics is provided by Peuquet's Triad Framework~\cite{Peuquet.TriadFramework.1994} which defines the interrelationship between \emph{where}, \emph{when}, and \emph{what}.
Several questions or tasks can be derived by placing these keywords either as the target or as a constraint (i.e. $when + where \rightarrow what$).
Andrienko and Andrienko~\cite{Andrienko.ExploratoryAnalysis.2006} extend this framework and derive a taxonomy of tasks for exploratory data analysis.
Since their tasks have been commonly used in the literature to facilitate analysis workflows~\cite{Gatalski.SpaceTimeCube.2004, Kristensson.CompareSpaceTimeCubeTo2D.2009, Demsar.SpaceTimeDensity.2010, Amini.InteractivityImpact2Dvs3D.2015, Filho.ImmersiveTableSpaceTimeCube.2020}, we use it as the foundation for our task definitions. 
\revise{The configuration of constraints and targets influences the applicability of operations, as indicated in \autoref{fig:infovis_pipeline}.
}

Tasks can be either \textbf{elementary} or \textbf{synoptic}: elementary tasks deal with individual references or datapoints, while synoptic tasks deal with sets of datapoints or groups.
A \textbf{behavior} generalizes notions like distributions, variations, and trends and is used to describe such a set.
%
Another distinction is made according to the task type: \textbf{lookup}, \textbf{comparison}, or \textbf{relation-seeking}.
When a reference or behavior is the target and is to be determined based on various constraints, we refer to a lookup task.
In comparison tasks, a reference or behavior is compared against another reference or value with the target of identifying a relation.
Relation-seeking tasks reverse the process and impose a relation as its constraint with the objective of determining references or behaviors that relate to these constraints.
Lastly, lookup and comparison tasks can be either \textbf{direct} or \textbf{inverse}.
In direct tasks, spatiotemporal information serves as the constraint to determine a thematic attribute as the target.
Inverse tasks take thematic attributes as constraints and look for spatiotemporal information.
\autoref{tab:tasks} provides an overview of possible task types and exemplary formulations, used as tasks in the expert study in~\autoref{sec:evaluation}.


\subsection{STC Operations}
\label{subsec:stc_ops}
In~\autoref{sec:related_work}, we outlined conceptual approaches of STC operations.
We base our operation methodology \revise{(and the icons\footnote{\vspace{-5mm}{B. Bach. Operation Icons, 2014. \href{https://aviz.fr/~bbach/spacetimecubes/}{https://aviz.fr/\textasciitilde bbach/spacetimecubes/}}})} on the framework provided by Bach et al.~\cite{Bach.SurveySpaceTimeCubeOperations.2016} as it provides a more exhaustive conceptualization compared to Li et al.~\cite{Li.TextSpaceTime.2020}. 
A STC operation can be defined as a transformation mapping that takes a STC and additional (optional) parameters to produce another STC. 
\begin{figure*}[t]
    \centering
    \includesvg[inkscapelatex=false, inkscapedpi=600, keepaspectratio, width=\linewidth]{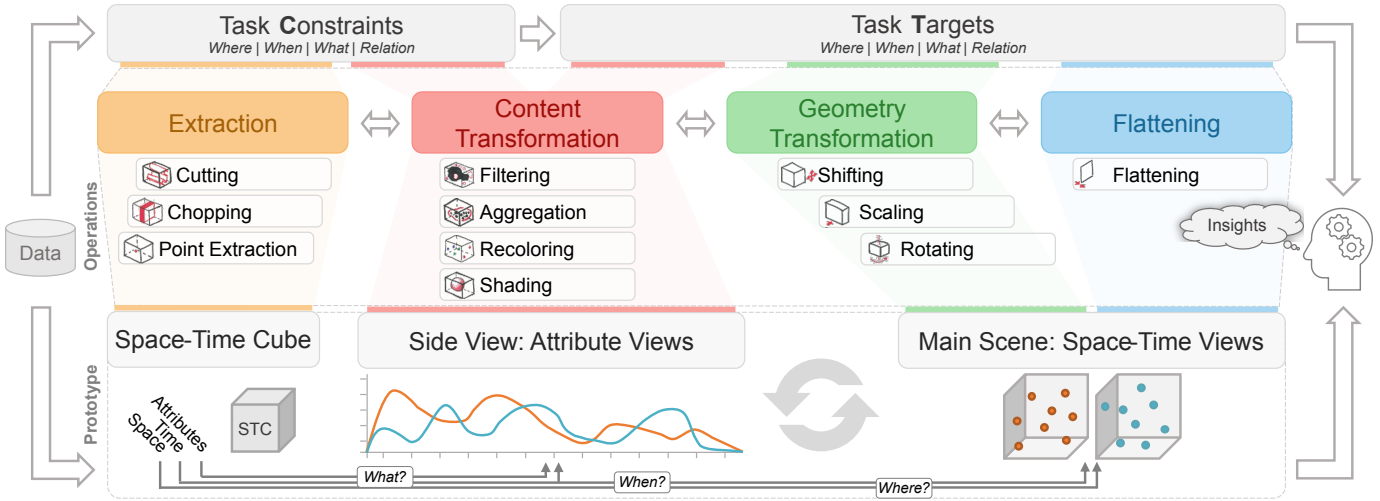}
    \vspace*{\figCapVSpace}
    \caption{\revise{Conceptual view of our \toolName~framework aligning analysis tasks (\autoref{subsec:st_tasks}), STC operations (\autoref{subsec:stc_ops}), and the prototype implementation (\autoref{sec:system}), highlighting how requirements associated with each operation are met through the user interface.}}
    \label{fig:infovis_pipeline}
    \vspace{\figVSpace}
\end{figure*}
The taxonomy of Bach et al.~\cite{Bach.SurveySpaceTimeCubeOperations.2016} classifies elementary STC operations into four categories:
\stcOperation{ex}{Extraction}, \stcOperation{ct}{Content Transformation}, \stcOperation{gt}{Geometry Transformation}, and \stcOperation{fl}{Flattening}.
The inner structure of the cube is hereby an important factor for choosing operations and is defined using the dimensions of \emph{density}, \emph{variability in positions}, \emph{variability in visual attributes}, and \emph{object lifespan}.
For spatiotemporal event data, the cube is typically rather sparsely populated with events often forming clusters.
Since events are discrete occurrences rather than continuous trajectories, they exhibit no variability in position or visual attributes over time. Additionally, unless a duration is explicitly provided, events can be considered to have an infinitesimally small lifespan, reducing them to points in space-time.
While other inner structures are possible and would affect design choices, we adapt our approach to the inner configuration most commonly found in spatiotemporal event datasets.
\revise{In the following, we outline each operation category from the original taxonomy~\cite{Bach.SurveySpaceTimeCubeOperations.2016}, and describe our integration choices and adaptations for task-driven spatiotemporal data analysis,  as shown in \autoref{fig:infovis_pipeline}.}

\textbf{Extraction} operations select a subset from the space-time volume. 
The \stcOperation{ex}{Point Extraction} operation extracts a single point, the \stcOperation{ex}{Drilling} operation a line, and the \stcOperation{ex}{Cutting} operation a 2D surface from the STC.
\stcOperation{ex}{Chopping} extracts a volume and is the only operation that again results in a STC.
These operations can be performed either in space or time, and have planar or non-planar (oblique, curvilinear) variants depending on the input parameters.
For sparsely dense cubes, operations such as Drilling or Cutting can often result in an empty extraction set, as the line or plane may not intersect any data points.
Instead, we consider extraction on either an \emph{elementary} or \emph{synoptic} level (akin to the tasks), where the former extracts single elements or references, and the latter a behavior.
This reduces the extraction operations we consider to a \stcOperation{ex}{Point Extraction} that retrieves individual datapoints, and a \stcOperation{ex}{Chopping} where a given behavior is extracted.
Chopping hence encompasses Drilling and Cutting as initially defined by Bach et al.~\cite{Bach.SurveySpaceTimeCubeOperations.2016}, where each operation extracts a subset of the data.
We redefine the \stcOperation{ex}{Cutting} operation as the act of placing a cut through the cube, effectively slicing the existing cube into two parts. 
\revise{In the context of analysis tasks, \stcOperation{ex}{Extraction} operations mainly serve the purpose of modeling the constraints by extracting selected behaviors.}

\textbf{Content Transformation} operations change the content of the STC without affecting the underlying geometry. 
These include \stcOperation{ct}{Filtering}, \stcOperation{ct}{Aggregation}, \stcOperation{ct}{Recoloring}, and \stcOperation{ct}{Shading}.
\stcOperation{ct}{Filtering} removes datapoints from a STC, \revise{and could in that sense also resemble an extraction operation.
Instead, we interpret this operation as a visual filtering or highlighting interaction, where the unselected datapoints are hidden.
A subsequent \stcOperation{ex}{Chopping} can then persist the selected data as a behavior in a new STC.
}
\stcOperation{ct}{Aggregation} combines multiple individual datapoints into a single, larger representation, simplifying the visualization of dense data. 
Clustering can also be viewed as a form of aggregation, where every datapoint is represented by its corresponding cluster label.
\stcOperation{ct}{Recoloring} changes the colors of datapoints in STCs according to a selected thematic attribute.
\stcOperation{ct}{Shading} is inherently implemented in a 3D environment, where light and shadow enhance depth perception and spatial understanding. 
We neglected various \stcOperation{ct}{Filling} (Time-, Space-, or Volume Interpolation)
and \stcOperation{ct}{Repositioning} (Stabilizing and Bundling) operations as they are not compatible with the inner cube structure of event-based datasets that we examined.
Depending on the intended usage, 
\revise{\stcOperation{ct}{Content Transformation} operations can both satisfy task constraints (i.e. filtering irrelevant data), as well as support identifying task targets (i.e. recoloring to compare spatial distributions of behaviors).}

\textbf{Geometry Transformation} incorporates operations that change the layout of the STC without altering the actual content. Such operations are \stcOperation{gt}{Shifting}, \stcOperation{gt}{Rotating}, and \stcOperation{gt}{Scaling}.
\stcOperation{gt}{Shifting} and \stcOperation{gt}{Rotating} comprise both camera movements within a scene and the repositioning of whole STC objects.
\stcOperation{gt}{Scaling} allows the resizing of both elements in an STC as well as resizing entire cubes.
Other Geometry Transformations such as \stcOperation{gt}{Bending} and \stcOperation{gt}{Unfolding} deform space-time objects, which is not really applicable to the inner structure of our cube consisting of elementary spatiotemporal events.
\revise{\stcOperation{gt}{Geometry Transformation} operations are beneficial for visual comparisons of STCs and their contents, and hence assist the identification of targets in analysis tasks.}

\textbf{Flattening} operations project the STC onto a lower-dimensional surface.
Time \stcOperation{fl}{Flattening} collapses the temporal axis to produce a 2D map, while space \stcOperation{fl}{Flattening} reduces the spatial axes to a timeline.
Although the abstract concept of oblique \stcOperation{fl}{Flattening}, where data is projected on a non-planar surface such as a trajectory plane, can be valuable for tracing events along a curved space-time path, we restrict ourselves to the possibilities of spatial and temporal \stcOperation{fl}{Flattening}.
\revise{When the target of an analysis task is exclusively either the spatial (\emph{where}) or temporal (\emph{when}), flattening the non-target dimension can reduce visual clutter and enhance the visibility of relevant patterns.}

\subsection{Prototype}
\label{sec:system}
\revise{Our framework shows how analytical tasks can be modeled using STC operations (see \autoref{fig:infovis_pipeline}). To make these operations practically accessible, we propose an interactive prototype that enables task-based exploration of spatiotemporal event data.}
Following the positive resonance of linked views in STC implementations~\cite{Kraak.SpaceTimeCubeGeoPerspective.2003, Kveladze.GVAtool.2015}, the prototype incorporates a \emph{Main Scene} visualizing STCs, accompanied by a \emph{Side View} (left) containing multiple linked views.
The individual components are described in detail below.
We conducted qualitative assessments with experts to iteratively refine our design decisions.

\subsubsection{Architecture}
The prototype follows a containerized architecture featuring an Angular frontend using WebGL (Three.js) for 3D hardware-accelerated rendering, a Python Flask backend for data processing and serving, and a PostgreSQL+PostGIS database.



\begin{figure}[b!]
    \centering
    \includegraphics[width=\linewidth]{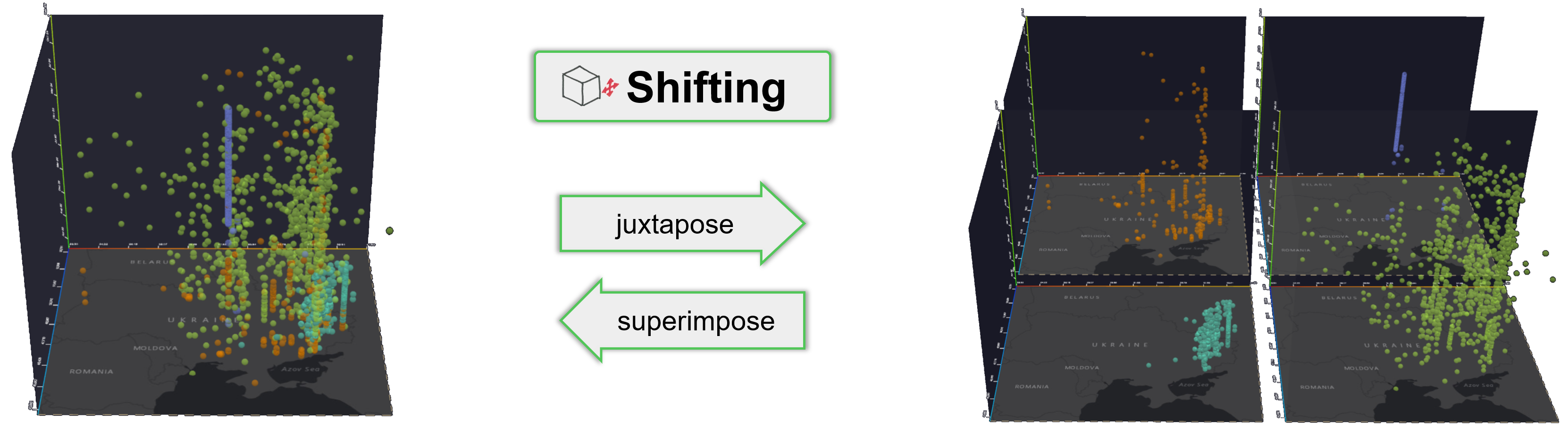}
    \vspace*{\figCapVSpace}
    \caption{Shifting permits a superimposed or juxtaposed layout of multiple behaviors (e.g. actors) obtained from repeated chopping.}
    \label{fig:juxt-superimp-layout}
    \vspace*{\figVSpace}
\end{figure}
\subsubsection{Main Scene: Space-Time Views}

\begin{figure*}[t]
    \centering
    \includegraphics[width=\linewidth]{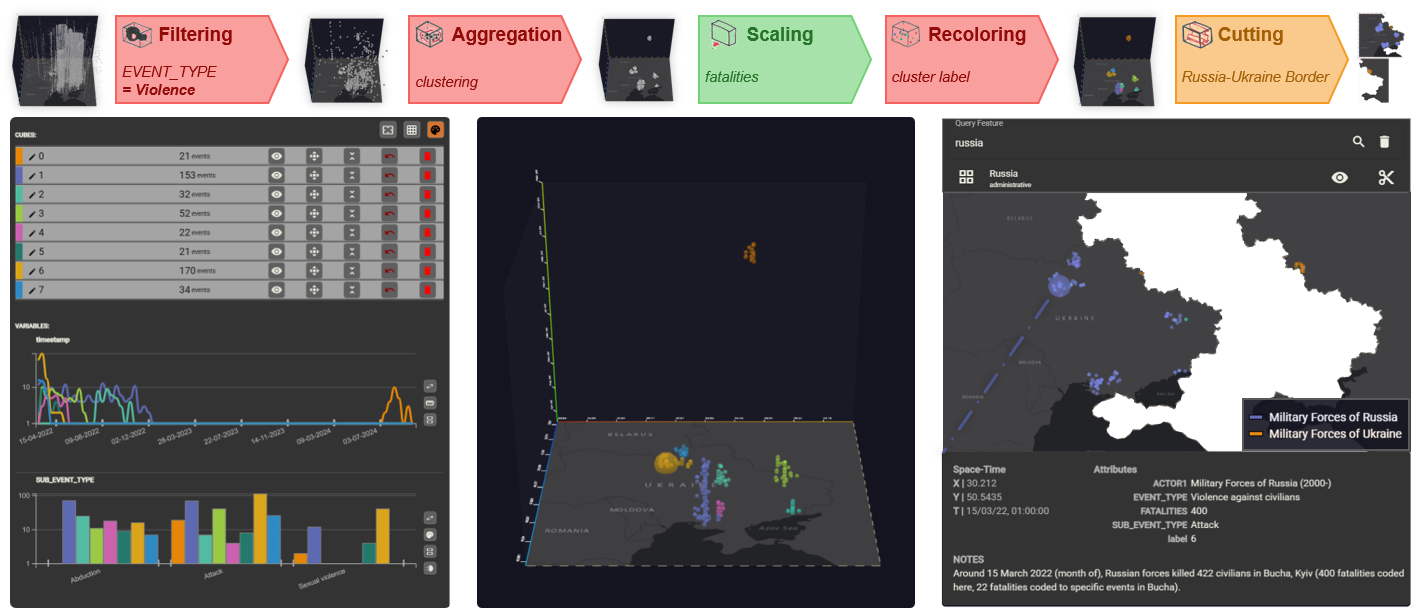}
    \vspace*{\figCapVSpace}
    \caption{\textbf{When and where did violence against civilians events occur?} After filtering, clustering reveals eight distinct groups, most concentrated early in the conflict. 
    Sizing according to \emph{fatalities} highlights more severe events such as the \emph{Bucha Massacre}.
    Recoloring according to \emph{actor} and performing a cutting along the Russia-Ukraine border shows that Russian acts occurred mainly on Ukrainian soil and vice versa.}
    \label{fig:case_study1}
    \vspace{\figVSpace}
\end{figure*}

The \emph{Main scene} is the canvas for displaying STCs, initially showing a single STC containing all events.
Every cube includes a geographical map as a base along the x and y-axis, while the temporal dimension extends upwards along the z-axis.
Every event is depicted as a sphere and positioned according to its spatial and temporal information. 
\stcOperation{gt}{Rotating}, \stcOperation{gt}{Shifting}, and \stcOperation{ct}{Shading} are inherently implemented as the camera can be panned and zoomed freely, and the scene is illuminated with a point light.
The scene is capable of handling multiple STCs as a core workflow concept consists of isolating behaviors in separate cubes.
Cubes can be freely translated within the scene, constituting a basic \stcOperation{gt}{Shifting} operation.
After receiving feedback on the initial prototype, a feature for a systematic layout of all cubes in the scene was requested.
To facilitate a visual comparison across space and time, we follow the notion of Gleicher et al.~\cite{Gleicher.comparison.2011} and provide means of arranging the cubes either in \emph{juxtaposing} or \emph{superimposed} configuration through a global, more sophisticated \stcOperation{gt}{Shifting} operation. 
Juxtaposition spans a squarified grid where all cubes are placed side by side, while superposition places the cubes at their respective initial location so that all visually appear as only one cube (see~\autoref{fig:juxt-superimp-layout}).
Upon hovering a sphere, the fully available information of a single datapoint is shown, performing a \stcOperation{ex}{Point Extraction} (see~\autoref{fig:case_study1}).
Based on feedback on our initial implementation, we added a helper line extending from the sphere to the bottom of the cube to enhance spatial awareness.


\subhead{Visual Variables}
The usage of visual variables allows to visually encode additional thematic attributes, hence facilitating visual multidimensional analysis. 
We followed the guidelines established by Kveladze et al.~\cite{Kveladze.stc_vis_variables.2019} for their usage within an STC environment.
Space and time are encoded using the x,y, and z \emph{position}.
\emph{Color hue} was effective for perceptually grouping categories, hence, we allow hue to encode categorical thematic attributes. We provide the \stcOperation{ct}{Recoloring} operation, assigning a distinct color to every available category.
\emph{Sizing} objects in a cube was deemed as effective for estimating quantities by scaling datapoints accordingly.
The \stcOperation{gt}{Resizing} operation can be performed with every numerical attribute of the dataset, scaling every sphere according to the selected attribute value.
Rotation, was found to be ineffective by Kveladze et al.~\cite{Kveladze.stc_vis_variables.2019}, and other variables such as shape or texture were not evaluated.
We did not consider them as their potential benefit is outweighed by the risk of overstimulation.

\begin{figure}[t!]
    \centering
    \includegraphics[width=\linewidth]{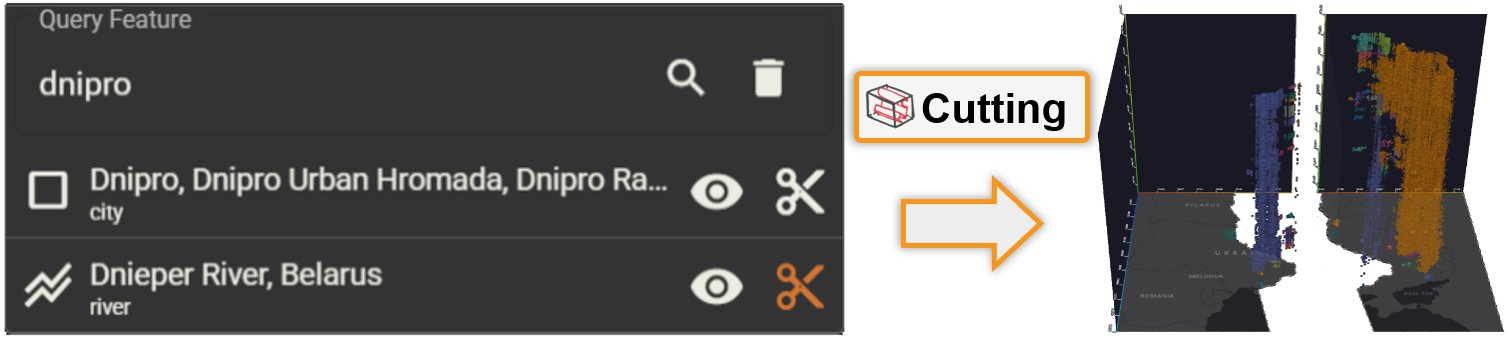}
    \vspace*{\figCapVSpace}
    \caption{Performing a Geospatial Query, leveraging the OSM Nominatim API to find the river \emph{Dnipro} and use it path for cube cutting.}
    \label{fig:query_feature}
    \vspace{\figVSpace}
\end{figure}
\begin{figure}[b!]
    \centering
    \includegraphics[width=\linewidth]{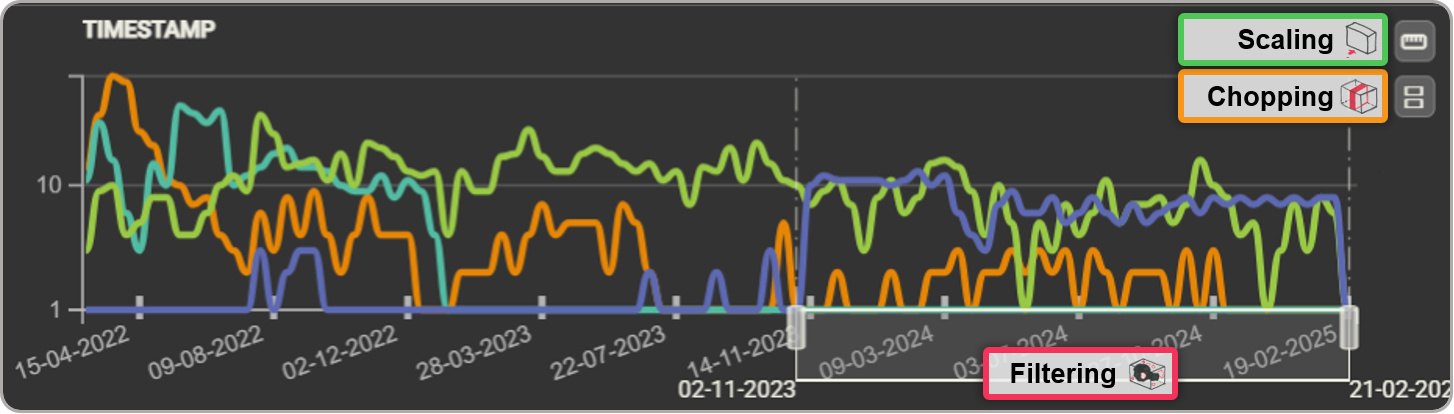}
    \vspace*{\figCapVSpace}
    \caption{Numeric attributes are summarized using KDE, each line represents a cube or behavior.
    Most \colorrect{c_naf}NAF events are recorded in 2022, while the \colorrect{c_marines}Ukraine Marines active interval starts in November 2024.  
    A filter can be set by brushing over the x-axis. 
    }
    \label{fig:numerical_dist_plot}
    \vspace*{\figVSpace}
\end{figure}

\subhead{Geospatial Query Feature}
Initially, spatial \stcOperation{ex}{Cutting} was implemented as moving a cutting plane along a cube's spatial axes. Although experts valued the operation, they noted that cuts aligned with latitude or longitude rarely yield meaningful insights. To add contextual relevance, we integrated a search bar powered by the OpenSteetMaps~\cite{osm:osm:17} Nominatim engine (see \autoref{fig:query_feature}) for querying geospatial features. This embeds geographic semantics into the workflow, enabling cuts along meaningful elements such as rivers or administrative boundaries.

\subhead{Clustering}
Clustering~\cite{Shi.stClusterSurvey.2019} can extract meaningful patterns, and Andrienko and Andrienko~\cite{Andrienko.ClusterAnalysisSTC.2009} demonstrated the potential of interactive clustering in an STC environment. 
We employ ST-DBSCAN~\cite{birant.stdbscan.2007}, while hyperparameters ($\epsilon 1$, $\epsilon 2$, $ min\_pts$) can be adapted interactively.
Every datapoint is then associated with a cluster label, resulting in a thematic attribute in the \emph{Side View} for \stcOperation{ct}{Aggregation}.

\subsubsection{Side View: Attribute Views}
The \emph{Side View} supports the interaction with multiple cubes through a cube list, attribute distributions, and a filter summary.

\begin{figure}[t!]
    \centering
    \includegraphics[width=\linewidth]{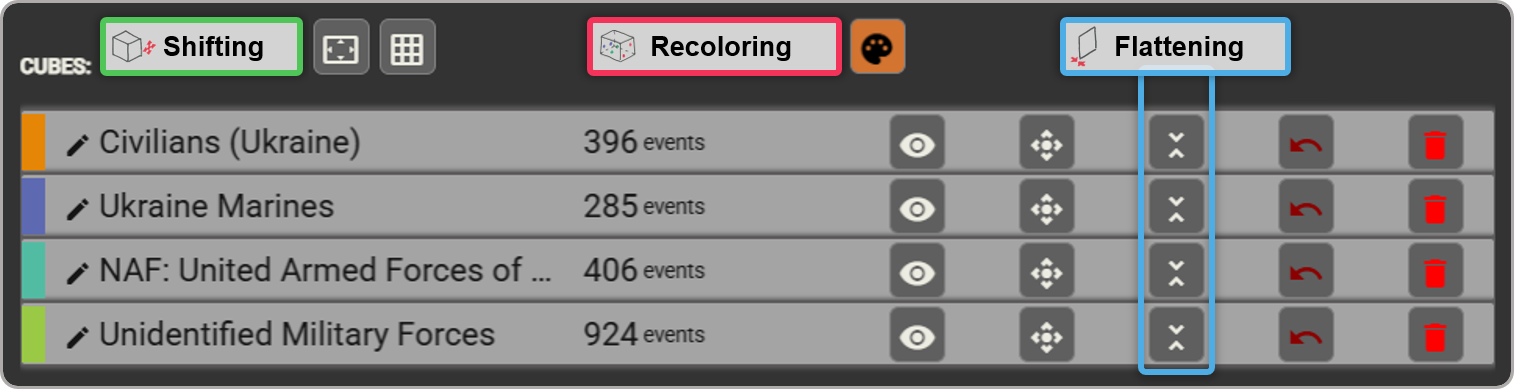}
    \vspace*{\figCapVSpace}
    \caption{The Cube List categorizes all created cubes or behaviors, with options for labeling, visibility toggling, shifting, and flattening.}
    \label{fig:cube_list}
    \vspace{\figVSpace}
\end{figure}

\subhead{Cube List}
To support the workflow of isolating behaviors in individual cubes, a cube list shows all created cubes or behaviors in a tabular view (see~\autoref{fig:cube_list}).
Every cube is associated with a label, allowing the user to associate the contained behavior with a textual description. 
Additional functions from this view include buttons to toggle its visibility, trigger a \stcOperation{gt}{Shifting} or \stcOperation{fl}{Flattening} operation, undo the last operation, or delete the cube.
The shortcut buttons to globally arrange all cubes according to a juxtaposed or superimposed configuration (seen in \autoref{fig:juxt-superimp-layout}), or \stcOperation{ct}{Recoloring} the data according to cubes are located at the top right of the cube list.

\subhead{Attribute Distributions}
When dealing with synoptic tasks, the analysis of behaviors comprising a set of datapoints is important.
Behaviors typically emerge from similarities across one or multiple spatial, temporal, or thematic attributes.
To summarize the behavioral distribution within STCs, a distribution plot for every thematic attribute is available.
We hereby fundamentally distinguish our design decisions between numerical and categorical attributes. 
While other attribute types exist that cannot be optimally represented (e.g., text, audio, video), we focus only on the most common numerical and categorical data types.

\begin{figure}[b!]
    \centering
    \includegraphics[width=\linewidth]{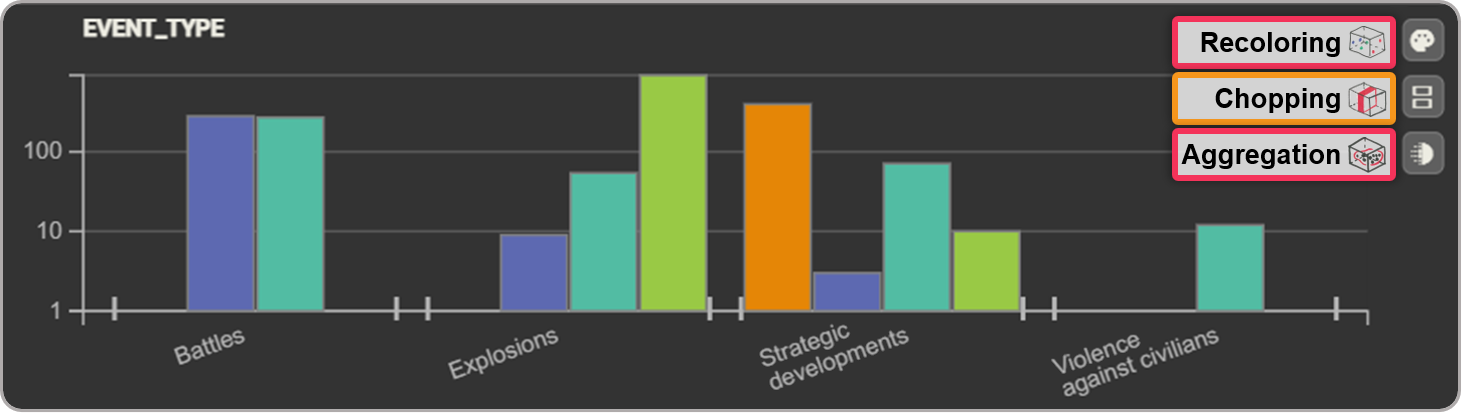}
    \vspace*{\figCapVSpace}
    \caption{Categorical attributes are summarized, e.g., through bar charts, with each bar representing a cube/behavior.
    \colorrect{c_unident}Unidentified Military Forces are responsible for most \emph{Explosions}. 
    \colorrect{c_naf}NAF are the only behavior to record \emph{Violence against civilians} events.}
    \label{fig:categorical_dist_plot}
    \vspace{\figVSpace}
\end{figure}

For \emph{numeric} thematic attributes, we use a kernel density estimation (see~\autoref{fig:numerical_dist_plot}).
With attribute values on the x-axis, the corresponding frequency can be read from the y-axis.
Every scene cube is summarized by a line, allowing a direct comparison of different behavior distributions.
Since size is considered an effective variable to estimate quantities~\cite{Kveladze.stc_vis_variables.2019}, we allow a \stcOperation{gt}{Resizing} of all spheres inside STCs according to the respective numerical attribute.
Multiple cubes according to numerical intervals can be created through a repeated \stcOperation{ex}{Chopping}, essentially discretizing the underlying data into $n$ sets of equal-width intervals.
When time is the numerical attribute, this constitutes a repeated temporal \stcOperation{ex}{Cutting} operation and allows the creation of \revise{periodically partitioned cubes}, i.e., one cube per month.

\emph{Categorical} 
We provide a frequency bar chart (see~\autoref{fig:categorical_dist_plot}) and, 
in case of multiple cubes, grouped bar charts.
This allows a direct comparison of quantities for every behavior across the categories.
While other visualizations, such as Mosaic Plots, are specifically designed for comparing relationships between multiple categorical attributes, we adhered to the concept of bar charts for visual consistency.
Color hue is an effective variable for perceptually grouping similar categories~\cite{Kveladze.stc_vis_variables.2019}, making them suitable candidates for a \stcOperation{ct}{Recoloring} operation.
Furthermore, \stcOperation{ct}{Aggregation} can also be performed on categorical attributes, replacing individual spheres with categorical isosurfaces through Marching Cubes~\cite{Lorensen.MarchingCubes.1987}.
To facilitate comparison across categories, we implemented a repeated \stcOperation{ex}{Chopping} operation of categorical attributes by creating subcubes for each category. 
This isolates every category as a behavior within a distinct cube, allowing comparison of their distribution across space, time, and other thematic attributes.


\subhead{Filtering}
Aside from visualizing the distributions of behaviors across the thematic attributes, the distribution plots serve as interactive plots to support the \stcOperation{ct}{Filtering} operation.
A brushing functionality permits the selection of an interval on a numeric attribute, which can subsequently be used to filter the cubes.
The bar charts register click events, allowing the user to filter for specific categories by clicking them. 
\revise{Events not matching the filter conditions are dynamically grayed out and reduced in size, realizing a visual \stcOperation{ct}{Filtering} operation, which can be persisted through a subsequent \stcOperation{ex}{Chopping} operation.}
The currently active filters are displayed in the filter summary at the top of the \emph{Side View}.
Every filter constraint is depicted as a textbox showing the attribute name and the corresponding criteria.
For numerical filters, the included value range is shown, and categorical filters feature a corresponding list.



\section{Evaluation}
\label{sec:evaluation}
We evaluate our framework through a case and an expert user study.

\subsection{Case Study}
\label{sec:case studies}
We use data from the Armed Conflict Location \& Event Data (ACLED)~\cite{raleigh.ACLED.political.2023} on the Russia-Ukrainian War (also called the Russo-Ukrainian War) from 01.02.2022 to 28.02.2025 (180k events, focusing primarily on Russia’s invasion of Ukraine) and apply a task-based approach to investigate key dynamics of conflicts: civilian victimization, weapon usage, and territorial shifts.

\subhead{When and where did violence against civilians events occur?}
This task investigates the temporal evolution and geographic distribution of events where \emph{Violence Against Civilians} was recorded (see~\autoref{fig:case_study1}).
To focus on relevant incidents, the analyst first applies a \stcOperation{ct}{Filtering} operation, isolating only events classified as \emph{Violence Against Civilians}. To gain an overview, the data is then \stcOperation{ct}{Aggregated} into clusters, revealing eight distinct groups. Most of these clusters are concentrated in the early stages of the war in 2022, with one exception occurring later in 2024.
To assess the severity of individual events, a \stcOperation{gt}{Scaling} operation is performed using the \emph{fatalities} attribute, confirming that the most fatal incidents took place during the initial phase of the conflict. 
A hover over the largest sphere identifies the Bucha Massacre as the deadliest event through a \stcOperation{ex}{Point Extraction}.
Next, the analyst uses \stcOperation{ct}{Recoloring} to determine the actors involved. The categorical distribution plot reveals that Russian forces were responsible for the majority of incidents.
Additionally, all early clusters are attributed to Russian forces, whereas the later cluster is linked to Ukrainian forces.
Finally, by performing a \stcOperation{ex}{Cutting} operation along the Russia-Ukraine border, the analyst confirms a geographic pattern: most Ukrainian-perpetrated violence occurred on Russian territory (in the Kursk oblast), whereas Russian-perpetrated violence predominantly took place on Ukrainian soil.

\subhead{Comparing Different Weapon Types}
This task investigates how weapon usage has developed over space and time (see~\autoref{fig:teaser}).
The \emph{event\_type} category \emph{Explosions/Remote violence} defines events where destructive weapon types have been used, and thus serves as a starting point for the exploration.
Upon \stcOperation{ct}{Filtering} the data to only include these events, the analyst can observe that the corresponding \emph{sub\_event\_types} are \emph{Air/drone strike}, \emph{Grenade}, \emph{Remote explosive}, and \emph{Shelling/artillery}.
To gain insights about the spatial distribution of these types, a \stcOperation{ct}{Recoloring} is performed.
For a more in-depth comparison of the types, the analyst performs repeated \stcOperation{ex}{Chopping} operations to isolate every \emph{sub\_event\_type} as a behavior in a separate cube.
Since the number of \emph{Grenade} (20) and \emph{Remote explosive} (1436) events are negligible compared to \emph{Air/drone strike} (39.071) and \emph{Shelling/artillery} (86.568) events, only the latter two are considered and kept as cubes.
Upon \stcOperation{gt}{Shifting}, the remaining two cubes are juxtaposed side by side.
To investigate the spatial distribution of these behaviors, \stcOperation{fl}{Flattening} along the temporal axis is performed. 
While the global distribution of both behaviors mainly follows the frontline, little variations can be observed, i.e. in the region of Belgorod.
In terms of temporal variance, a view on the attribute distribution plot reveals a diverging pattern.
Before June 2023, \emph{Shelling/artillery} was the dominating \emph{sub\_event\_type}. 
Since then, the number of \emph{Air/drone strikes} has been steadily increasing and took the lead since June 2024.

\begin{figure*}[t]
    \centering
    \includegraphics[width=\linewidth]{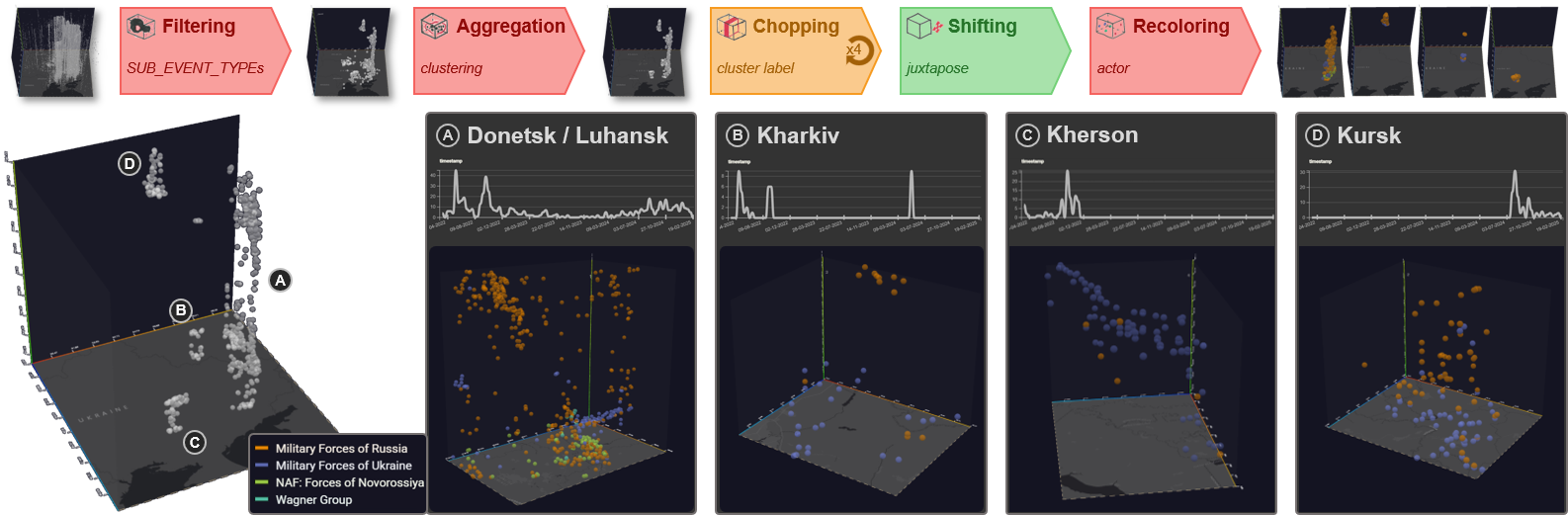}
    \vspace*{\figCapVSpace}
    \caption{\textbf{Analyzing Patterns of Territorial Gains and Losses}. After filtering for relevant \emph{sub\_event\_types}, clustering reveals four distinct regions of territorial exchanges, namely Donbas (Donetsk / Luhansk), Kharkiv, Kherson, and Kursk regions. By chopping and juxtaposing them, they can be examined closer with respect to involved \emph{actors} and temporal development. }
    \label{fig:case_study2}
    \vspace*{\figVSpace}
\end{figure*}




\subhead{Analyzing Patterns of Territorial Gains and Losses}
This task determines locations of shifts of territorial ownership, characterized by multiple transfers over time (see~\autoref{fig:case_study2}).
Through \stcOperation{ct}{Filtering}, the analyst selects the relevant \emph{sub\_event\_types}: \emph{Government regains territory}, \emph{Non-state actor overtakes territory}, and \emph{Non-violent transfer of territory}.
To identify spatial clusters, \stcOperation{ct}{Aggregation} is performed through clustering while neglecting the temporal component, resulting in four distinct spatial clusters.
Each cluster is then isolated for further analysis using repeated \stcOperation{ex}{Chopping} and \stcOperation{gt}{Shifting} to juxtapose them.
A \stcOperation{ct}{Recoloring} operation by \emph{actor} then reveals the forces involved in each territorial shift.
The largest cluster, in the Donbas region, exhibits three phases: (1) In May 2022, Russian forces with the support of the paramilitary NAF and Wagner Group, overtook territories from Ukraine. (2) In September 2022, Ukraine regained areas mainly in the northern regions. (3) By July 2024, Russian forces advanced further southward in Donetsk.
A second cluster in the Kharkiv region follows a similar timeline within a closer spatial extent: Russian occupation, the recapture by Ukraine in September 2022 and its repeated loss to Russia in June 2024.
A third cluster in Kherson reveals a smaller-scale exchange, where Russian forces made some gains in March 2022 before Ukraine pushed back in October 2022.
The final cluster, in the Kursk region, corresponds to a Ukrainian incursion in August 2024, followed by a slow Russian offensive.

\subsection{Expert User Study}
\label{sec:userstudy}



We conducted pair analytics sessions~\cite{pairanalytics} with eight expert participants (\usr{1 -- 2} PostDoc level, \usr{3 -- 8} PhD level)\revise{, all professional data scientists with 4--10 years of experience with spatiotemporal datasets, with varying expertise regarding 3D visualizations and STCs.} 
The study assesses the usability of performing analysis tasks with the embedded operations and identifies those utilized.

\subsubsection{Procedure}
After signing a consent letter and filling out a demographic questionnaire, each participant was made familiar with our system through a demo video.
The assessment comprised two major parts: a task-based phase (with the ten tasks outlined in~\autoref{tab:tasks}), followed by a free exploration phase without any constraints.
\revise{A task was considered as completed as soon as the participant provided a justified answer using the prototype.}
We utilized three ACLED~\cite{raleigh.ACLED.political.2023} datasets.
Central Africa (65k) in the introductory video, events in the Middle East (114k) for the task-based, and the Russia-Ukrainian War (180k) for the free exploration phase.
The study was concluded with a closing questionnaire where participants reported insights from both phases and ranking our system through the ten questions of the System Usability Score (SUS)~\cite{Brooke.SUStest.1996}.

\subsubsection{Results}
When asked about their initial expectations beforehand, the participants identified the detection of spatiotemporal hotspots or clusters, the analysis of complex, possibly repeating, multidimensional patterns, as well as the developments over time.
Aside from providing an overview, the prototype shall also be able to provide details on demand~\cite{Shneiderman.Mantra.1996}.
Basic filtering and highlighting capabilities, as well as more complex features, such as querying geospatial entities, were also stated.

\subhead{Task-based}
All participants heavily relied on initial \stcOperation{ct}{Filtering} and \stcOperation{ex}{Chopping} to meet the given constraints and reduce the number of events. 
They were the first applied operations for every task with the exception of \tsk{1}, where most participants first used \stcOperation{ex}{Cutting} to isolate the geographical region and extract the point from the resulting cube. 
\usr{1 -- 5} applied \stcOperation{fl}{Flattening} whenever temporal information was not required, relying on a 2D representation instead of a 3D STC.
\tsk{4} (inverse synoptic lookup) posed the greatest difficulty for the participants, where \usr{2, 3} were not able to complete the task.
The other participants either used the attribute distribution of the timestamp in the side view (\usr{4, 6, 8}) or looked for a gap of events in the cube (\usr{1, 5, 7}) to identify a time interval with no explosions.
For \tsk{8} (inverse synoptic comparison of spatial distributions), \usr{2, 4, 5, 8} relied on the filtered cube paired with \stcOperation{ct}{Recoloring}, whereas \usr{1, 3, 6, 7} created individual cubes for each pattern by \stcOperation{ex}{Chopping} and juxtaposed these using \stcOperation{gt}{Shifting}.

\subhead{Free Exploration}
In the free exploration phase, \usr{1, 2, 6} were interested in the events with the highest fatalities, and were able to locate them in the cube using \stcOperation{ct}{Filtering} by brushing the corresponding distribution plot.
\usr{4, 5} specifically searched and found events they had previously encountered in news reports, using both \stcOperation{ct}{Filtering} and \stcOperation{ex}{Cutting} (Bucha Massacre, Kerch Bridge explosion, Kursk offensive).
\usr{1, 2, 3, 7} identified the shift in explosion events outlined in \autoref{fig:teaser} by repeated \stcOperation{ex}{Chopping} and subsequent \stcOperation{gt}{Shifting} operations, while \usr{3, 4} investigated the spatial and temporal distribution of violence against civilian events using \stcOperation{ct}{Aggregation} and \stcOperation{ct}{Recoloring} (akin to~\autoref{fig:case_study1}).

\begin{figure}[t]
    \centering
    \hspace*{-2mm}
    \includegraphics[width=1.02\linewidth]{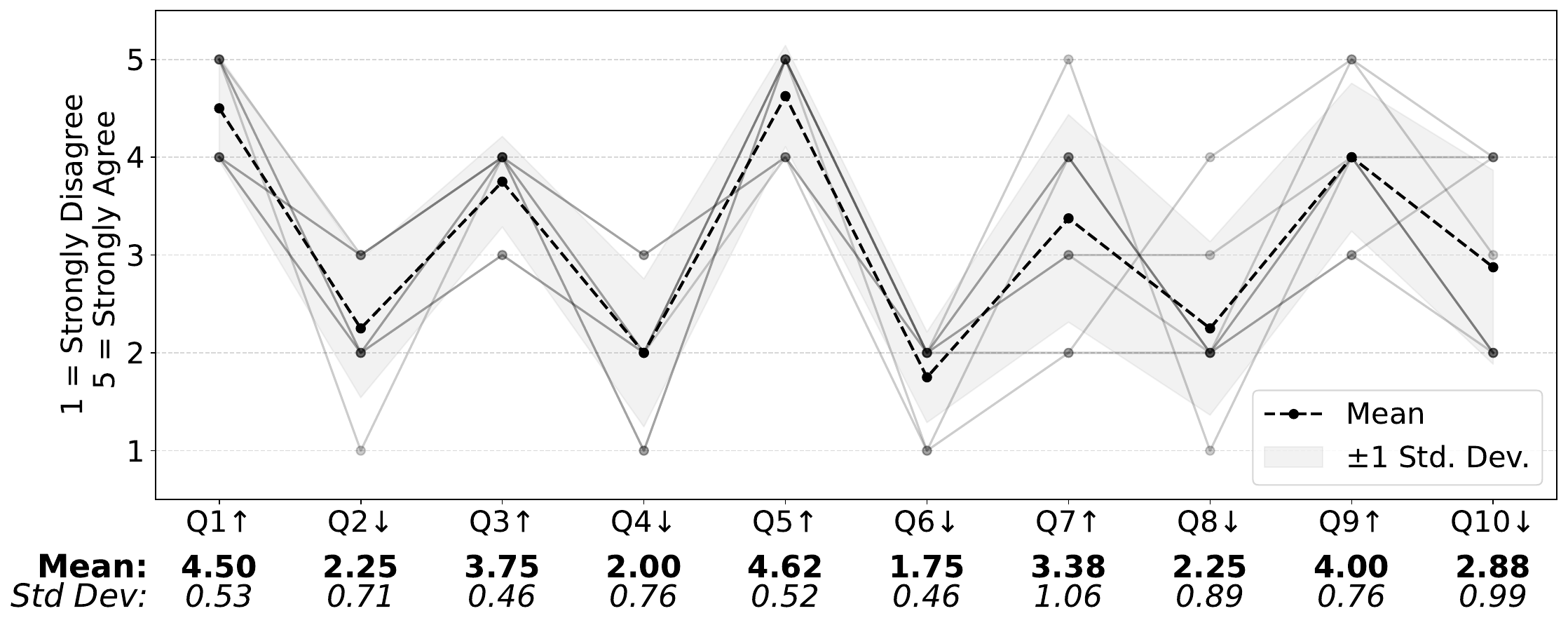}
    \vspace*{\figCapVSpace}
    \caption{Results for the SUS by question, each line representing one study participant. A zigzag pattern (odd questions are to be maximized, even questions minimized), is visible for Q1-Q6. However, Q7, Q8, and Q10 show deviations with highest SD.}
    \label{fig:sus_results}
    \vspace{\figVSpace}
\end{figure}

\subhead{SUS}
The computation of the total SUS score resulted in a value of 72.81, scoring above the average score of 68 and yielding a \emph{Good} result (B-) according to~\cite{Lewis.SusScoring.2018}.
\autoref{fig:sus_results} shows the individual responses of each question.
Since the even-numbered questions of the SUS are reverse-coded, 
the ideal result would yield a zigzag pattern.

\subhead{Feedback}
As the SUS scores indicate (especially Q1 and Q5), the overall feedback was positive.
\usr{2} expressed that the initial introduction of the tool and operations was important to understand the entire workspace, and that confidence while using the tool grew over time.
Especially the spatial \stcOperation{ex}{Cutting} (to isolate or compare geographic regions) or the repeated \stcOperation{ex}{Chopping} followed by \stcOperation{gt}{Shifting} (e.g. to juxtapose categorical attributes) was appreciated by the participants.
Some participants criticized that extracting individual events was cumbersome, which is especially important for completing elementary tasks.
\usr{2} and \usr{3} stated that the selection of individual days through the brushing of the temporal distribution plot to be impractical, and \usr{8} noted that the overlapping of spheres in close spatiotemporal proximity made \stcOperation{ex}{Point Extractions} by hovering difficult.
When comparing distributions of multiple cubes or behaviors, \usr{3, 6} stated that the categorical distribution plot can suffer from visual clutter due to an increased number of very small rectangles, making it difficult to discern patterns.
Additional features requested included panning reset and default viewpoint return (\usr{5}), reordering the cubes in the cube list as well as the juxtaposed layout configuration (\usr{7}), seeing the filter options applied to a cube (\usr{8}), and using the query bar for temporal and thematic attribute information (\usr{3} and \usr{6}).


\section{Discussion}
\label{sec:discussion}
In this chapter, we present findings from the expert user study,
and outline limitations and directions for future work.

\subsection{Findings and Lessons Learned}
\label{sec:findings}

The results from the expert user study indicate that the prototype effectively allows analysts to solve spatiotemporal tasks with only two out of 80 tasks \emph{not} being successfully completed and therefore skipped by the participants (2.5\%).
The overall SUS score of 72.81 suggests good usability. Participants particularly appreciated the integration of operations within the system (Q5).
Especially the \stcOperation{ex}{Cutting} operation as well as the repeated \stcOperation{ex}{Chopping} was greatly appreciated to facilitate a comparison between behaviors based on spatial, temporal, or thematic constraints.
However, responses to Q7 and Q10 suggest a learning curve, implying that familiarity with the system is necessary for efficient use. 
While this is expected for specialized analytical tools, future refinements could further improve intuitiveness.

\subhead{Extended Prototype}
\revise{
Based on the user feedback and additional observations, we extended the prototype with the following features:
To enhance temporal linking capabilities between the \emph{Side View} and the cubes in the \emph{Main Scene}, we display a movable translucent plane in each cube when the Temporal Attribute Distribution Plot is hovered, indicating the hovered temporal position.
The currently set temporal filter range is further highlighted with red boundary struts, marking the lower and upper bounds of the range.
To facilitate the exploration of recurrent temporal patterns, the temporal information can be transformed by retaining only specific components of the timestamp.
The available transformation modes include yearly, monthly, weekly, and daily layouts, supporting the investigation of seasonal trends, monthly cycles, weekly routines, or daily fluctuations.
Other minor improvements include data import from csv files, an option to reset the view, the possibility to reorder behaviors in the cube list, and customization options for visual parameters such as sphere sizes, time-axis scaling, and map style.}

\subhead{Further Study Implications}
Some users reported that functionalities relevant for elementary tasks (e.g. \stcOperation{ct}{Filtering} on a single day, \stcOperation{ex}{Point Extraction} for occluded spheres) were slightly impractical. This is in line with research about STCs claiming that simpler tasks are easier to solve with a 2D visualization~\cite{Kristensson.CompareSpaceTimeCubeTo2D.2009, Amini.InteractivityImpact2Dvs3D.2015}.
Conversely, STC environments are more suited when sets or behaviors have to be analyzed, and our framework follows this notion with its synoptic-oriented approach.
Of all given tasks, \tsk{4} presented the greatest challenge for most participants, where a temporal interval with no occurrences of a given attribute (\emph{Explosions}) was to be determined.
Participants either rotated the cube to focus on the time axis or looked at the distribution over time in the attribute view.
The size of the spheres in the cube imposed an issue, as larger spheres possibly obscured the interval in the STC visualization. 
\revise{The short duration of the interval (24th-30th Nov) likely further complicated the task, where above-mentioned prototype extensions (sphere sizes, time-axis scaling) could reduce occlusion problems.}
For \tsk{8}, we observed two main strategies:
One group used \stcOperation{ex}{Chopping} followed by \stcOperation{gt}{Shifting} to have a juxtaposed, side-by-side view of the two behaviors that should be compared.
Another solution approach was to keep both behaviors in one cube and distinguish the events by \stcOperation{ct}{Recoloring}.
Both approaches led to a similar result but provided different viewpoints on the same data, emphasizing our framework's flexibility.

During the free exploration phase, participants largely adhered to a structured approach, from which we distill key insights about user workflows. The first step was consistently \stcOperation{ct}{Filtering} and subsequent \stcOperation{ex}{Chopping}, used to satisfy given constraints and reduce visual clutter.
\stcOperation{ct}{Aggregation} can be used to gain an overview of broader trends, filter out noise, and identify interesting behaviors, which can subsequently be isolated by creating separate cubes using repeated \stcOperation{ex}{Chopping}.
To facilitate comparison of these behaviors, two distinct strategies emerged: some juxtaposed behaviors using \stcOperation{gt}{Shifting} to place them side by side, while others opted for superposition and applied \stcOperation{ct}{Recoloring} to distinguish behaviors within a single cube. 
Finally, participants leveraged both spatial insights in the 3D \emph{Main Scene} and attribute distributions in the \emph{Side View} to identify patterns and validate their findings.

\subhead{Operations}
The study indicates that our framework based on Bach et al.'s operations~\cite{Bach.SurveySpaceTimeCubeOperations.2016} offers a solid basis for task-driven analysis.
While we do not question the validity of their taxonomy, some conceptual inconsistencies emerged in practice.
\stcOperation{ct}{Filtering} is defined as a content transformation, but the general act of filtering resembles an extraction operation that we instead realized through \stcOperation{ex}{Chopping}.
The implemented \stcOperation{ct}{Filtering} follows the \emph{Select} method by Brehmer and Munzner~\cite{brmer.vistasks.2013}, which is not unambiguously represented in Bach et al.'s taxonomy.
\stcOperation{gt}{Scaling} is classified as a geometry transformation, although changing the sphere sizes according to a data attribute can be seen as a content transformation.
\stcOperation{fl}{Flattening}, while intended as a projection onto a lower dimensional surface, can also be interpreted as an \stcOperation{ct}{Aggregation} operation, especially when coupled with hexagonal binning techniques or heatmaps.
Because the initial taxonomy was primarily developed to classify visualizations rather than analytical workflows, it does not account for practical meta operations. Features such as hiding cubes or undoing the last operations fall outside the established categories, even though they were frequently used by experts during the study.



Overall, the results from both the case and the expert user studies demonstrate how different visualizations on the same tasks can be created with the operations embedded in our framework and how this can effectively support decision-making.
We publish the framework as \href{https://osf.io/qgr4x/overview?view_only=48a6e22e60574460a47ef24102f3c78a}{open-source}, believing it can have an impact on practical domains and provide a solid foundation for further research.

\subsection{Limitations and Future Work}
\label{sec:limitations_future_work}
Given the complexities of geo analytics, some \emph{limitations} exist.

\subhead{Comprehensiveness of Operations}
Bach et al.~\cite{Bach.SurveySpaceTimeCubeOperations.2016} derive planar, curvilinear, and oblique variants for all Extraction and Flattening operations.
Aside from \stcOperation{ex}{Cutting} where we realize curvilinear space cuts, we only provide the most basic planar variants.
Oblique extractions are difficult to comprehend and their practicality remains to be investigated.
When extending the framework to additional spatial datatypes such as trajectories, operations that we have discarded (i.e. \stcOperation{gt}{Bending} or \stcOperation{ct}{Filling}) would gain relevance.


\subhead{Limited Datatypes}
The current implementation of the prototype focuses on analyzing event-based datasets, and the derivation of operations has been tailored to such datasets.
Previous STC implementations \cite{Demsar.SpaceTimeDensity.2010, Kveladze.GVAtool.2015, Filho.ImmersiveTableSpaceTimeCube.2020} focused on trajectory or movements as their main datatype.
Extending the analysis to line-, and polygon-based geographic entities would further broaden the capabilities of the prototype.
Aside from the spatial component, handling other thematic data types, such as unstructured text, video, or audio data remains an open challenge. 

\subhead{3D Rendering}
Our approach inherits common limitations of 3D visualizations, such as occlusion, perspective distortion, and depth ambiguity.
Either \stcOperation{ct}{Filtering}, \stcOperation{ct}{Aggregation} or the extraction of behaviors through \stcOperation{ex}{Chopping} can reduce overplotting and alleviate occlusion issues.
\stcOperation{fl}{Flattening} allows to reduce the cube to a 2D representation where perspective distortions can be minimized.


\vspace{.5em}
\noindent Based on our findings and the feedback from our domain experts, we identify opportunities for \emph{future work}. 

\subhead{Technical Enhancements}
Existing methods such as space-time density estimation~\cite{Demsar.SpaceTimeDensity.2010} or 3D pencil icons~\cite{Tominski.3DPencils.2005} could be integrated.
Clustering currently uses ST-DBSCAN, while e.g. COD-CLARANS~\cite{Tung.codclarans.2001} would respect obstacles, and MDST-DBSCAN~\cite{Choi.mdstdbscan.2021} was developed to integrate thematic attributes.

\subhead{User Studies}
A more in-depth, larger user study could provide quantitative evidence about the suitability of different operations (or different combinations to the same effect) for a given task, also compared to 2D maps, timelines, and traditional GIS applications. 
More broadly, our framework can serve as a uniform benchmark for the recreation and comparability of previous perceptual STC studies~\cite{Kristensson.CompareSpaceTimeCubeTo2D.2009, Kjellin.2Dvs3D.2010, Kveladze.GVAtool.2015, Amini.InteractivityImpact2Dvs3D.2015, Kveladze.stc_vis_variables.2019}.

\subhead{Immersiveness} 
Enabling analysts to fully immerse themselves within the space-time cube operation framework (and not a 3D world projected on a 2D screen) would provide novel interaction modalities and facilitate collaborative analysis.
Future work could therefore explore the integration into immersive environments such as virtual reality (VR) or augmented reality (AR).
Initial research exists~\cite{Filho.ImmersiveTableSpaceTimeCube.2020, Fouche.ImmersiveHyperSpaceTimeCube.2023, Wagner.TaxiVisVR.2024}, but an interesting avenue could be to investigate interaction modalities.


\section{Conclusion}
\label{sec:conclusion}

In this work, we introduced \toolName, a framework that
bridges theoretical models of STC operations with practical analysis workflows by modeling tasks through a sequence of elementary operations.
An open-source VA prototype that implements these operations enables a structured and interactive analysis of spatiotemporal event data, facilitating both exploratory and confirmatory analysis. 
A case study on the Russo-Ukrainian war demonstrates how \toolName\ can uncover meaningful spatiotemporal patterns, while an expert user study validates the framework’s effectiveness in supporting real-world analysis tasks. The study further provided insights into user workflows, revealing common operation sequences and strategies and reinforcing the potential of structured STC operations to enhance analytical processes in the future.



\bibliographystyle{abbrv-doi-hyperref}

\bibliography{bibliography}





\end{document}


\maketitle
\onecolumn

\renewcommand{\thefigure}{A.\arabic{figure}}
\setcounter{figure}{0}

\section{Further UI Elements and Transformations}

\begin{figure}[!h]
    \centering
    \includegraphics[width=0.3\linewidth]{figures/appendix/hover_plane.png}
    \hspace{0.2\linewidth}
    \includegraphics[width=0.3\linewidth]{figures/appendix/filter_box.png}
    \caption{Temporal indicators inside the STCs. Left: A plane that is shown when hovering over the Temporal Attribute Distribution Plot. Right: If a temporal filter is set, the extent of the filter range is indicated by red boundary struts. }
    \label{fig:time-interaction-indicators}
\end{figure}

\begin{figure}[!htb]
    \centering
    \includegraphics[width=\linewidth]{figures/appendix/temporal_transformations.png}
    \caption{Different temporal transformations, showing cyclical patterns of demonstrations in Israel. While the yearly and monthly alignments do not seem to exhibit any patterns, the weekly alignment shows that most demonstrations take place on Saturdays. }
    \label{fig:time-transformations}
\end{figure}

\clearpage
\section{Operations}

\begin{figure}[!htb]
    \centering
    \includegraphics[width=\linewidth]{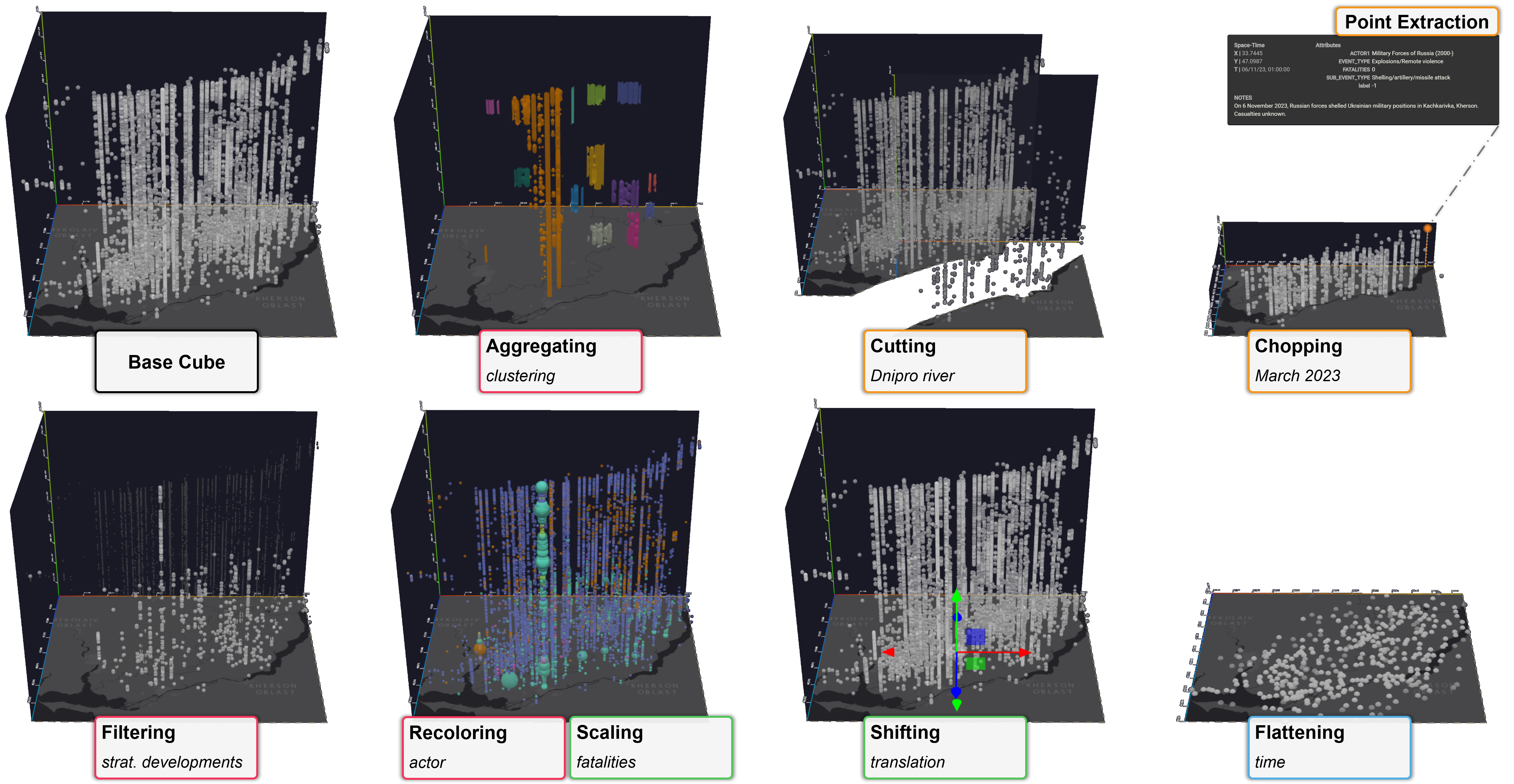}
    \caption{Overview of the main categories of STC operations implemented in \toolName, each with a specific example. }
    \label{fig:stc-ops-overview}
    \vspace{\figVSpace}
\end{figure}

\clearpage

\section{System Usability Scale (SUS)}

\begin{table}[h!]
\centering
\begin{tabular}{llcc}
\toprule
\textbf{ID} & \textbf{Question} & \textbf{Mean} & \textbf{Std. Dev.} \\ 
\midrule
\textbf{Q1} ($\uparrow$) & \textit{I think that I would like to use this system frequently.}                                   & 4.50  & 0.53 \\
\textbf{Q2} ($\downarrow$) & \textit{I found the system unnecessarily complex.}                                                  &  2.25 & 0.71 \\
\textbf{Q3} ($\uparrow$) & \textit{I thought the system was easy to use.}                                                      &  3.75 & 0.46 \\
\textbf{Q4} ($\downarrow$) & \textit{I think that I would need the support of a technical person to be able to use this system.} &  2.00  & 0.76 \\
\textbf{Q5} ($\uparrow$) & \textit{I found the various functions in this system were well integrated.}                         &  4.62 & 0.52 \\
\textbf{Q6} ($\downarrow$) & \textit{I thought there was too much inconsistency in this system.}                                 & 1.75 & 0.46 \\
\textbf{Q7} ($\uparrow$) & \textit{I would imagine that most people would learn to use this system very quickly.}              &  3.38 & 1.06 \\
\textbf{Q8} ($\downarrow$) & \textit{I found the system very cumbersome to use.  }                                               &  2.25 & 0.89 \\
\textbf{Q9} ($\uparrow$) & \textit{I felt very confident using the system.}                                                    &  4.00  & 0.76 \\
\textbf{Q10} ($\downarrow$) & \textit{I needed to learn a lot of things before I could get going with this system.}               &  2.88 & 0.99 \\
\bottomrule
\end{tabular}
\caption{Questions, Mean Answers, and Standard Deviations of the System Usability Scale (SUS). The results were gathererd using a 5-Point Likert Scale (1 = Strongly Disgree, 5 = Strongly Agree).}
\end{table}

The final SUS score was calculated according to the formula:

\begin{equation}
    SUS = 2.5~( 20 + \sum(Q1, Q3, Q5, Q7, Q9) - \sum(Q2, Q4, Q6, Q8, Q10) ) = 72.81
\end{equation}